\documentclass{article}

\usepackage{preprint_style,times}

\usepackage{microtype}
\usepackage{graphicx}
\usepackage{subcaption}
\usepackage{booktabs}
\usepackage{tikz}
\usetikzlibrary{arrows.meta,positioning,fit}
\usepackage{placeins}
\usepackage{afterpage}
\usepackage{hyperref}
\usepackage{url}

\usepackage{amsmath}
\usepackage{amssymb}
\usepackage{mathtools}
\usepackage{amsthm}

\usepackage[capitalize,noabbrev]{cleveref}

\theoremstyle{plain}

\theoremstyle{definition}

\theoremstyle{remark}

\usepackage[textsize=tiny]{todonotes}

\title{UNAAGI: Atom-Level Diffusion for Generating Non-Canonical Amino Acid Substitutions}

\author{
Han Tang \\
Department of Computer Science (DIKU) \\
University of Copenhagen \\
\texttt{ht@di.ku.dk} \\
\And
Alexander Tong \\
AITHYRA \\
\texttt{atong@aithyra.at} \\
\And
Wouter Boomsma \\
Department of Computer Science (DIKU) \\
University of Copenhagen \\
\texttt{wb@di.ku.dk}
}

\makeatletter
\def\section{\@startsection{section}{1}{\z@}{-1.2ex plus -0.5ex minus -.2ex}%
  {0.9ex plus 0.3ex minus 0.2ex}{\large\sc\raggedright}}
\def\subsection{\@startsection{subsection}{2}{\z@}{-1.08ex plus -0.5ex minus -.2ex}%
  {0.48ex plus .2ex}{\normalsize\sc\raggedright}}
\def\subsubsection{\@startsection{subsubsection}{3}{\z@}{-0.9ex plus -0.5ex minus -.2ex}%
  {0.3ex plus .2ex}{\normalsize\sc\raggedright}}
\makeatother
\prepfinalcopy  % arXiv preprint: reveal authors, drop the review line numbers

\begin{document}

\maketitle

% The base style sets \lhead inside \@maketitle, so this must follow it.
\lhead{Preprint. Under review.}

\begin{abstract}
A central challenge in structural biology is identifying beneficial amino acid substitutions, a problem that underpins both mutational effect prediction and protein engineering. Recent inverse-folding models, trained to reconstruct sequences from structure, have shown considerable promise for identifying functional mutations. However, current approaches are constrained to designing sequences composed exclusively of canonical amino acids (CAAs). Non-canonical amino acids (NCAAs) offer greater chemical diversity and are frequently used for protein engineering in vivo, yet they remain largely inaccessible to current variant effect prediction methods.

To address this gap, we introduce \textbf{UNAAGI}, a diffusion-based generative model that reconstructs residue identities from atomic-level structure using an E(3)-equivariant framework. By modeling side chains in full atomic detail rather than as discrete tokens, UNAAGI enables the exploration of canonical and non-canonical amino acid substitutions within a unified generative paradigm. We evaluate UNAAGI zero-shot, without mutational-effect supervision or task-specific fine-tuning, on experimental mutational-effect benchmarks and demonstrate substantially improved performance on NCAA substitutions relative to existing methods. 
\end{abstract}

\section{Introduction}

\afterpage{%
\begin{figure*}[!t]
    \centering
    \includegraphics[width=0.8\textwidth]{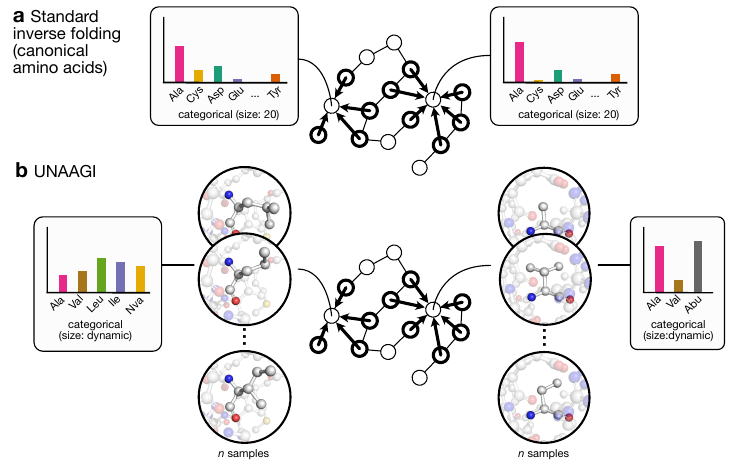}
    \caption{
        Illustration of UNAAGI compared with a standard inverse-folding model.
        \textbf{(a)} Inverse-folding models parameterize the output distribution
        as a categorical distribution over a fixed-size vocabulary.
        \textbf{(b)} UNAAGI generates amino acids directly in chemical space.
        A finite set of samples defines an empirical distribution over the
        context-dependent region of chemical space, typically including both
        canonical amino acids (CAAs) and non-canonical amino acids (NCAAs).
    }
    \label{vis_intro}
\end{figure*}
}

Proteins perform diverse biological functions through three-dimensional structures formed by linear chains of amino acids. Although amino acid side chains can, in principle, span broad chemical diversity, naturally occurring proteins primarily use a set of 20 genetically encoded canonical amino acids (CAAs). This set is thought to be evolutionarily optimized for functional completeness and broad chemical coverage \citep{ilardo2015extraordinarily, doig2017frozen, ilardo2019adaptive}. Nevertheless, non-canonical amino acids (NCAAs) provide side-chain chemistries absent from the standard set, including enhanced metal coordination, tailored electrostatics, and novel catalytic properties. Incorporating NCAAs therefore expands the functional repertoire of proteins beyond evolutionary constraints, opening new possibilities in synthetic biology and therapeutics \citep{link2003non, chin2017expanding, rogers2018nonproteinogenic}.

In recent years, machine learning---particularly deep learning---has achieved remarkable success in protein research, advancing tasks such as structure prediction, design, and mutational effect estimation \citep{Jumper2021, Dauparas2022, Abramson2024}.
%Among these tasks, the \textit{inverse folding problem} aims to recover a plausible amino acid sequence that folds into a given target structure. 
%Before the rise of deep learning approaches like ProteinMPNN, traditional methods such as Rosetta addressed this problem by estimating the energy of a sequence given a structure and identifying the sequence that minimizes this energy function \citep{alford2017rosetta}. 
Variant effect prediction models are often trained in an unsupervised manner by predicting amino acid propensities conditioned on sequence \citep{riesselman2018deep} or structural \citep{Dauparas2022} context. These methods generally represent amino acids with a discrete distribution over the 20 canonical residues, learned from large evolutionary databases. Although such models have shown impressive zero-shot performance across various protein properties, their discrete representation prevents generalization beyond the canonical amino acid vocabulary, creating a fundamental limitation in the current modeling paradigm.

A central question is whether inverse-folding models can be generalized beyond canonical amino acids. This problem presents two challenges: (i) the vast chemical space of NCAAs makes categorical modeling impractical, and (ii) the predictive capabilities of inverse-folding models arise from evolutionary constraints encoded in millions of naturally occurring protein sequences, a source of information that is not available for NCAAs.

% In this paper, we explore a hypothesis which could provide a potential solution to both challenges: by modeling amino acids generatively in full atomic detail, rather than as a set of discrete tokens, we can obtain meaningful transfer from the natural to the non-canonical amino acids. While it is clear that we cannot expect prediction capabilities to extend far out-of-domain, we hypothesize that it is possible to predict the substitution feasibility for non-canonical amino acids which do not deviate dramatically from their canonical counterparts.

% In this paper, we test whether atom-level generative modeling can provide a route beyond fixed amino-acid vocabularies. Rather than representing residues as discrete tokens, we model side chains in full atomic detail, allowing the model to learn local geometric and chemical constraints that may transfer from canonical amino acids to chemically proximal NCAAs. This formulation bypasses the need to define a fixed-size dictionary while leveraging abundant canonical amino acid data to design beyond it. Although such transfer is unlikely to cover the full diversity of NCAA chemical space, it may enable structure-conditioned assessment of chemically proximal NCAA substitutions whose side-chain chemistry remains close to that of canonical residues.

In this paper, we test whether atom-level generative modeling can provide a route beyond fixed amino-acid vocabularies. The key idea is that local geometric and chemical constraints learned from canonical amino acids may transfer to chemically proximal NCAAs. This formulation addresses both challenges: it avoids defining a fixed NCAA vocabulary while leveraging abundant canonical amino acid data to support design beyond the canonical set. Although such transfer is unlikely to cover the full diversity of NCAA chemical space, it may enable structure-conditioned assessment of substitutions whose side-chain chemistry remains close to that of canonical residues.

To test this hypothesis, we propose a generative framework for residue-level sequence design based on atom-level side-chain generation through equivariant molecular diffusion. We term our approach \textbf{U}ncanonical \textbf{N}ovel \textbf{A}mino \textbf{A}cid \textbf{G}enerative \textbf{I}nference (\textbf{UNAAGI}). By modeling side-chain atomic coordinates directly, UNAAGI implicitly infers residue identity and can propose plausible non-canonical substitutions. Figure~\ref{vis_intro} illustrates the central idea.

To evaluate the generalizability of UNAAGI, we first trained models while holding out selected canonical amino acids from the training set, testing whether UNAAGI could recover unseen amino-acid chemistries, particularly at structural sites where the held-out residue is experimentally favored. We then evaluated UNAAGI trained on the complete dataset using experimentally measured mutational effects \citep{rogers2018nonproteinogenic, proteingym}, comparing its generative likelihoods with observed substitution outcomes. UNAAGI substantially improves predictive performance for non-canonical substitutions relative to existing methods, while maintaining consistent performance across canonical and non-canonical residues.

Specifically, our contributions are as follows:

\begin{itemize}
    \item We introduce UNAAGI, an E(3)-equivariant diffusion framework for atom-level side-chain generation and residue identity inference.
    \item We evaluate controlled generalization beyond the training vocabulary by holding out selected canonical amino acids and testing whether UNAAGI can recover them as unseen chemistries.
    \item We demonstrate that UNAAGI can generalize to chemically proximal non-canonical amino acids, enabling structure-conditioned assessment of NCAA substitutions.
    \item We evaluate UNAAGI in zero-shot variant effect prediction, without mutational-effect supervision or task-specific fine-tuning, and demonstrate improved performance on experimental NCAA substitution benchmarks for the identified chemically proximal subclass.
    % \item We discuss the connection between NCAA-aware protein engineering and structure-based drug design, suggesting a path toward unified atom-level generative modeling across both settings.
\end{itemize}

\section{Related Work}
\label{related_work}

%Direct comparison to baseline methods is challenging. The most comparable works are \citet{Geylan2025} and \citet{Lee2025.07.31.667780}. \citet{Geylan2025} aims to generate peptide sequences containing NCAA substitutions using a token-level, language-based approach. However, their work lacks direct experimental validation. We therefore include PepINVENT \citep{Geylan2025} and NCFlow \citep{Lee2025.07.31.667780} as the most relevant prior methods to UNAAGI, while also broadening our perspective to other related areas, including molecular diffusion and mutational effect prediction. In this way, we position our contribution within a broader context.

Our approach connects prior work in molecular diffusion, variant effect prediction, and peptide design.

\subsection{Molecular Diffusion}

\citet{hoogeboom2022equivariantdiffusionmoleculegeneration} introduced E(3)-equivariant diffusion frameworks for molecular generation, enabling the direct generation of molecular structures in 3D space through a diffusion process. Subsequent works, including \citet{peng2023moldiffaddressingatombondinconsistency} and \citet{vignac2023midimixedgraph3d}, incorporated additional inductive biases from molecular bonding structures, leading to improved molecular consistency and sample quality. More recently, \citet{le2024navigating} conducted a systematic comparison of several molecular diffusion frameworks, examining how different architectural and modeling choices affect generative performance and identifying combinations of design choices that yield strong performance across molecular generation tasks.

Beyond unconditional molecular generation, an important application is the conditional generation of compounds that bind to specific protein targets, commonly known as structure-based drug design (SBDD). \citet{Schneuing2024} extended the framework of \citet{hoogeboom2022equivariantdiffusionmoleculegeneration} to conditional ligand generation in 3D space, while \citet{guan20233dequivariantdiffusiontargetaware} proposed related models for target-aware ligand sampling. More recently, \citet{schneuing2025multidomain} introduced models that sample ligands with variable atom counts while incorporating protein-pocket side-chain conformations during generation.

\subsection{Mutational Effect Prediction}

Mutational effect prediction aims to estimate how amino acid substitutions affect a protein property of interest, such as stability, affinity, or function. Several branches of machine learning models have demonstrated sensitivity to this problem, including pure sequence-based models, structure-to-sequence (inverse-folding) models, and hybrids of the two

Statistical models of multiple sequence alignments remain strong baselines when many sequences are available for a given protein family. Notable examples are the DeepSequence VAE model and the EVE model \citep{riesselman2018deep, Frazer2021DiseaseVP}. Language-based models such as ESM \citep{meier2021language, lin2023evolutionary} obviate the need for alignments,
leveraging conservation and coevolutionary signals learned from large protein sequence datasets. Although sequence-based models can predict mutational effects in many contexts, their performance declines when evolutionary signals are sparse, as in shallow MSAs, antibodies, or de novo designed peptides.

Structure-based approaches replace the sequence context of a mutation with structural context. Early models considered the local environment around one amino acid at a time \citep{boomsma2017spherical,torng20173d}; this idea was later generalized to entire sequences in inverse-folding models \citep{ingraham2019generative,Dauparas2022,pmlr-v162-hsu22a}.
%, including inverse folding models \citep{Dauparas2022, pmlr-v162-hsu22a}, reconstruct sequences from given backbone structures under the assumption that the generated sequences will fold exclusively into the specified structure. 
Although primarily developed for protein design, these models are also useful for mutational effect prediction \citep{frellsen2025zero}. %Recent work \citep{su2023saprot} has combined sequence and structure information to further improve performance.

Many existing all-atom protein models still operate in two stages: they first predict a canonical amino acid identity and then generate its corresponding atomic conformation. UNAAGI similarly models one residue at a time, conditioned on its surrounding structural environment. The crucial difference is that UNAAGI generates the side chain directly at the atomic level, without first assigning it to a predefined amino acid class. This formulation allows the model to represent side-chain chemistries beyond the canonical amino-acid alphabet and thereby extend mutational effect prediction to non-canonical amino acids.

%Our approach similarly considers one amino acid at a time, with the crucial difference that
%Although our approach does not strictly align with the traditional inverse folding problem—since we focus on recovering a single residue at a time—it nonetheless demonstrates awareness of the mutational landscape and can be evaluated under similar benchmarks. 
%Crucially, whereas prior methods are restricted to the natural amino acid space, 
% UNAAGI extends mutational effect prediction to include non-canonical amino acids.

\subsection{Modeling Non-canonical Amino Acid Substitutions}

%As discussed in the previous section, proposing NCAA substitutions to proteins and predicting their mutational effects is a promising direction for structural biology, but remains under-explored within the machine learning for structural biology community. Related works, including \citet{Geylan2025}, \citet{Li2025.05.19.654846}, and \citet{Lee2025.07.31.667780}, have also investigated proposing NCAA substitutions or designing peptides containing NCAAs.

PepINVENT \citep{Geylan2025} is a transformer-based language model trained on atomic-level peptide representations using CHUCKLES \citep{Siani1994CHUCKLES}, which converts amino acid residue tokens into SMILES strings. This representation enables the model to generate amino acid SMILES strings at masked positions. The strategy supports generalization to non-canonical amino acids, but the original study does not validate the resulting sample probabilities against experimental measurements. We therefore include PepINVENT as a baseline.

RareFold \citep{Li2025.05.19.654846} and NCFlow \citep{Lee2025.07.31.667780} are based on AlphaFold-style architectures. RareFold introduces new tokens for NCAAs and focuses on structure prediction for proteins containing NCAA substitutions and on the design of NCAA-containing peptides. The study reports experimentally validated peptide designs, but the model itself does not produce in silico affinity or mutational-effect scores, making direct comparison on NCAA mutational effect prediction infeasible. NCFlow adopts a flow-matching framework and evaluates mutational effect prediction on the same benchmark as ours, enabling direct comparison (Section~\ref{experiments}).

\section{UNAAGI}
\label{sec:UNAAGI}

We introduce UNAAGI, a diffusion-based model that bridges molecular generation and protein mutational effect prediction by constructing amino acid side chains atom by atom, thereby extending the mutational landscape beyond the canonical amino acid space. The model learns patterns in
%By learning from 
the chemistry of canonical residues and uses them to generalize to chemically proximal non-canonical amino acids.
%UNAAGI can generalize and interpolate to non-canonical amino acids. 
We next outline equivariant graph diffusion and then proceed by describing the model architecture, graph construction, and sampling procedure.

\subsection{Multimodal Diffusion for Molecular Graphs}

We summarise the standard denoising-diffusion formulation, and the equivariance
definitions used below, in Appendix~\ref{app:prelim}.

Molecular data are inherently \textit{multimodal}: atomic coordinates are continuous, whereas atom types, bond types, and other chemical features are discrete. Diffusion over discrete variables can be defined through categorical transition kernels, which are discrete analogues of Gaussian noise in the forward process \citep{hoogeboom2021argmaxflowsmultinomialdiffusion, austin2023structureddenoisingdiffusionmodels}. Recent molecular diffusion models combine continuous and categorical processes to jointly model these heterogeneous modalities \citep{peng2023moldiffaddressingatombondinconsistency, vignac2023midimixedgraph3d, guan20233dequivariantdiffusiontargetaware, le2024navigating}.

The forward noise process for continuous and discrete modalities can be expressed as
\begin{equation}
\begin{aligned}
q(\mathbf{x}_t \mid \mathbf{x}_0) &= \mathcal{N}\!\left(\mathbf{x}_t \mid \sqrt{\bar{\alpha}_t} \mathbf{x}_0,\ (1 - \bar{\alpha}_t)\mathbf{I}\right), \\
q(\mathbf{c}_t \mid \mathbf{c}_0) &= \mathcal{C}\!\left(\mathbf{c}_t \mid \bar{\alpha}_t \mathbf{c}_0 + (1 - \bar{\alpha}_t) \tilde{\mathbf{c}}\right),
\end{aligned}
\end{equation}
where \( \bar{\alpha}_t = \prod_{k=1}^{t} (1 - \beta_k) \in (0, 1) \) is the cumulative noise schedule. For discrete features, \( \tilde{\mathbf{c}} \) denotes the categorical prior (e.g., uniform or empirical distribution). Following D3PM \citep{austin2023structureddenoisingdiffusionmodels}, we adopt a \textit{mask diffusion} scheme, which gradually converts labels into an absorbing state during the forward process.

In our setting, we perturb both atomic coordinates \( \mathbf{X} \) and atom-wise categorical features \( \mathbf{H} \). The features in \( \mathbf{H} \) include atom types, formal charges, hybridization states, aromaticity, ring membership, and atomic degree. Each modality is diffused independently using either Gaussian or categorical noise. This chemically-aware perturbation strategy is inspired by \citet{le2024navigating}, who showed that injecting chemically meaningful noise improves sample quality and stability.

The training objective is the standard diffusion ELBO, stated in full in
Appendix~\ref{app:prelim}. It is typically optimized by predicting either the clean data \( \hat{\mathbf{x}}_0 \), the noise \( \boldsymbol{\epsilon} \), or categorical logits, depending on the parameterization \citep{ho2020denoisingdiffusionprobabilisticmodels, austin2023structureddenoisingdiffusionmodels}. We adopt the \( \hat{\mathbf{x}}_0 \)-parameterization to predict the clean data directly. The loss combines mean-squared error for atomic coordinates with cross-entropy for categorical features.
\subsection{E(3)-Equivariant Graph Neural Network}

Molecular structures are inherently three-dimensional and symmetric under Euclidean transformations such as rotation and translation. To respect these symmetries, we incorporate \textit{equivariance} as an inductive bias. The relevant symmetry group is $\mathrm{E}(3)$ (rotations, reflections and
translations); the formal equivariance conditions are given in
Appendix~\ref{app:prelim}.

We adopt an E(3)-equivariant graph neural network, closely following the EQGAT-diff model of \citet{le2024navigating}, to learn the score function over atom-level graphs. Each node represents an atom and carries scalar features \( h_i \) (e.g., atom type and charge) and vector features \( \mathbf{v}_i \in \mathbb{R}^3 \), while edges encode bond types and interatomic distances. Further architectural and training details are provided in Appendix~\ref{suppl1}.

\subsection{Graph Topology and Virtual Node Padding}

Unlike traditional structure-based drug design (SBDD), in which generative models sample ligands conditioned on a protein pocket, our task focuses on reconstructing amino acid side chains. Each generated side chain is covalently bonded to a fixed protein backbone, imposing local structural constraints around the masked region. We therefore fix the positions and types of the backbone atoms for the residue under reconstruction, anchoring generation to the known structural context.

A further challenge is unified sampling across canonical and non-canonical amino acids, whose side chains contain different numbers of atoms. In standard SBDD models, the atom count must be fixed before diffusion because the graph topology is predetermined.
We address this challenge using a virtual-node strategy inspired by DrugFlow \citep{schneuing2025multidomain}. Unlike empirical size-sampling schemes \citep{hoogeboom2022equivariantdiffusionmoleculegeneration, Schneuing2024, guan20233dequivariantdiffusiontargetaware} or separate size-estimation networks \citep{Igashov2024}, this approach is fully end to end and allows the model to learn a distribution over variable side-chain sizes.

Virtual nodes are assigned a special atom type \texttt{NOATOM} and are disconnected from all other atoms, with edges labeled \texttt{NOBOND}. During training, a random number of virtual nodes \( n_{\text{virt}} \sim U(0, N_{\max}^{\text{train}}) \) are added to the side chain. Following \citet{schneuing2025multidomain}, these nodes are initialized at the side-chain center of mass, allowing the model to learn which nodes correspond to real atoms. At sampling time, the model denoises a graph initialized with \( N_{\max}^{\text{sample}} \) nodes. Nodes denoised to the virtual type are removed post hoc, allowing variable-sized side chains to be generated within a unified diffusion framework.

\section{Experiments}
\label{experiments}
We now describe the experimental procedure for UNAAGI, including training details, evaluation methods, and comparisons against baselines.

\subsection{Dataset Composition and Preprocessing}
\label{data_prep}

We construct a curated training dataset by combining data from multiple sources:

\begin{itemize}
\item \textbf{Canonical protein structures.} We use a filtered set of protein structures from the Protein Data Bank (PDB), with full dataset construction details provided in Appendix~\ref{train_details}. For each structure, we extract every residue together with its local structural environment, defined as all surrounding residues whose centers of mass lie within 10~\AA~of the target residue.

\item \textbf{Non-canonical amino acids.} To expose the model to chemical structures beyond the canonical amino-acid alphabet, we incorporate non-canonical amino acids (NCAAs) from \citet{ilardo2019adaptive} and SwissSidechain \citep{Gfeller2013SwissSidechain}. These datasets contain isolated amino-acid structures without associated protein environments. During training, the four backbone atoms (N, C$_\alpha$, C, and O) are kept fixed, while noise is applied only to the side-chain atoms.

\item \textbf{Protein--ligand complexes.} We include protein--ligand complexes from PDBBind \citep{wang2005pdbbind} to expose the model to a broader range of local chemical interactions and ligand-like molecular structures. Following the structure-based drug design setup of \citet{le2024navigating}, we keep the protein-pocket atoms fixed and apply noise to all ligand atoms.

\item \textbf{PDB structures containing NCAAs.} We additionally collect protein structures from the PDB that contain non-canonical amino acids, providing examples in which NCAA side chains occur in explicit protein environments. Details of the collection and filtering procedure are provided in Appendix~\ref{train_details}.

\end{itemize}

To assess the contribution of each supplementary data source, we perform ablation experiments in which individual components are removed from the training set and evaluate the resulting changes in model performance. Further details are provided in Section~\ref{ablation}.

% We train on a filtered set of protein structures from the Protein Data Bank (PDB); full dataset details are provided in Appendix~\ref{train_details}. For each structure, we extract every residue together with its local environment, defined as all surrounding residues whose centers of mass lie within 10~\AA~of the target residue.

% In addition to canonical amino acids, we include non-canonical amino acids (NCAAs) to improve the model's ability to generalize to novel chemical environments and reduce overfitting. We use NCAA datasets from \citet{ilardo2019adaptive} and SwissSidechain \citep{Gfeller2013SwissSidechain}, although these contain isolated NCAAs without associated protein context. To expose the model to a broader range of molecular interactions, we also augment the training set with protein--ligand complexes from PDBBind \citep{wang2005pdbbind}. Graph construction differs across data sources:
% \begin{itemize}
%    \item For isolated NCAAs, the four backbone atoms (N, C$_\alpha$, C, and O) are fixed during training, and only the side-chain atoms are diffused.
%    \item For PDBBind, we fix the protein-pocket atoms and apply noise to all ligand atoms, following the SBDD procedure of \citet{le2024navigating}.
% \end{itemize}

% We additionally fine-tune the model on PDB structures containing NCAAs, as described in Appendix~\ref{train_details}.

% We conduct an ablation study on the supplementary sources of training data to evaluate their individual effects on model performance. Details are provided in Section~\ref{ablation}.
\subsection{Accelerated Sampling and Residue Identification}
\label{sampling}
Given a fixed local protein environment, UNAAGI generates variable-size side-chain molecular graphs by jointly predicting atomic coordinates, atom features, and bond types. After virtual nodes are removed, generated side chains are assigned to known canonical or non-canonical amino acids by matching their labeled molecular graphs to reference structures using graph isomorphism \citep{Weisfeiler:1968ve}. We use the NetworkX 3.5 implementation of \texttt{is\_isomor\allowbreak{}phic}, which is based on the VF2 algorithm \citep{cordella2001improved}.

Downstream evaluation requires many independent samples at each residue position to estimate the model-induced distribution over amino acid identities (Section~\ref{dms_exp}). Standard ancestral DDPM sampling requires 500 denoising steps per sample and is therefore computationally expensive. At inference time, we instead apply DPM-Solver++ \citep{lu2025dpm} to the continuous coordinates and continuous-time Markov chain updates \citep{campbell2022continuoustimeframeworkdiscrete} to the categorical atom and bond features. Together, these methods reduce the number of denoising steps from 500 to 20 per sample.

% The model architecture, diffusion framework, and loss functions are detailed in Section~\ref{sec:UNAAGI}.
\subsection{Rediscovering Canonical Amino Acids from Held-out Models}
\label{subsection_rediscover}
Large-scale and diverse experimental ground truth for NCAA mutational effects is unavailable, preventing direct evaluation of whether a generative model designs them appropriately. We therefore construct a surrogate task by removing one canonical amino acid entirely from training. The model must then treat the held-out residue as unseen chemistry and reconstruct it atom by atom, analogous to the setting encountered for a true NCAA. Using ProteinGym measurements, we test whether the model can (i) generate the held-out residue and (ii) place it at positions where it is favored. We train four separate held-out models, each excluding one amino acid both as a reconstruction target and from the surrounding structural context to prevent data leakage.

% \subsubsection{Recovery Rate of the held-out amino acid: Can UNAAGI generate the amino acids it has never seen?}

We compare each held-out model with a model trained on all 20 canonical amino acids. For evaluation, we randomly select 25 assays from ProteinGym \citep{proteingym} and separately hold out Ser, Ile, Phe, and Val. As shown in Table~\ref{tab:heldout-generation}, the held-out models recover residues that never appear in their training data, both at native sites and at positions where the substitution is stabilizing. The models also exhibit selectivity, generating the held-out residue more frequently at favorable positions (native and stabilizing sites combined) than at harmful positions.

\begin{table}[t]
\centering
\small
\setlength{\tabcolsep}{4pt}
\begin{tabular}{lcccccccc}
\toprule
& \multicolumn{1}{c}{Rebuilds native} & \multicolumn{3}{c}{Designs at favorable sites} & \multicolumn{4}{c}{Prefers favorable over harmful sites} \\
\cmidrule(lr){2-2}\cmidrule(lr){3-5}\cmidrule(lr){6-9}
Excluded & without it & without it & full model & \% kept & favorable & harmful & fold & $p$ \\
residue & (rate) & (rate) & (rate) & vs.\ full & sites (rate) & sites (rate) & change & \\
\midrule
VAL & 0.990 & 0.921 & 0.972 & 95\% & 0.928 & 0.843 & $1.10\times$ & $3.2\mathrm{e}{-6}$ \\
SER & 1.000 & 0.896 & 0.977 & 92\% & 0.908 & 0.720 & $1.26\times$ & $1.2\mathrm{e}{-18}$ \\
PHE & 0.929 & 0.674 & 0.896 & 75\% & 0.687 & 0.589 & $1.16\times$ & $4.2\mathrm{e}{-4}$ \\
ILE & 0.885 & 0.357 & 0.863 & 41\% & 0.407 & 0.295 & $1.38\times$ & $1.1\mathrm{e}{-4}$ \\
\bottomrule
\end{tabular}
\caption{\textbf{The model generates held-out residues with a systematic difficulty gradient.} Each value is the fraction of sites at which a model trained \emph{without} the excluded residue nonetheless generates it (``without it''), compared with the model trained on all 20 canonical amino acids (``full model''). \textbf{Rebuilds native} is reconstruction ($\mathrm{wt}=X$); \textbf{designs at favorable sites} is generalization ($\mathrm{wt}\neq X$, with mutation to $X$ stabilizing according to deep mutational scanning); \textbf{prefers favorable over harmful sites} is selectivity, and ``fold change'' is the ratio of those two rates. Per-residue site counts, Fisher exact-test $p$-values and pooled Wilson intervals are given in Appendix~\ref{AA_recon}.}
\label{tab:heldout-generation}
\end{table}
% The detailed results for this experiment could be found in Supplementary ~\ref{AA_recon}
\subsection{Evaluation on Deep Mutational Scanning (DMS)}
\label{dms_exp}
Given a fixed local protein environment, UNAAGI generates diverse side-chain identities and conformations. We interpret the resulting frequencies over residue identities as an empirical estimate of the model-induced distribution at each position.

To evaluate predictive performance, we compare this learned distribution with experimental deep mutational scanning (DMS) data.\footnote{ProteinGym is also used in Section~\ref{subsection_rediscover}, but there only to identify favorable and harmful substitution sites for the held-out-residue analysis. The present section directly evaluates mutational-effect prediction using DMS benchmarks containing both canonical and non-canonical amino-acid substitutions.} For each site in a DMS benchmark, we perform 1,000 sampling iterations and record the frequency of each sampled amino acid.\footnote{We assign unsampled amino acids a pseudocount of 1 (Laplace add-one smoothing), equivalent to the posterior mean of the sampling distribution under a uniform Dirichlet(1) prior. This bounds the log-likelihood ratio when a substitution is never sampled. In Appendix~\ref{pseudocount_robustness}, we assess robustness by excluding unsampled amino acids from the evaluation.} Although larger sample sizes would provide smoother frequency estimates, we use 1,000 iterations as a trade-off between statistical stability and computational cost. We repeat each experiment five times and report error bars to quantify sampling variability.

From the sampling frequencies, we estimate the probabilities assigned to the mutant and wild-type residues. We define the mutation score as the difference between their negative log-likelihoods:

$$
\Delta \mathrm{NLL}
=
-\log P(\text{mutant})
+
\log P(\text{wild-type}),
$$

where \(P(\text{mutant})\) and \(P(\text{wild-type})\) denote the estimated probabilities of the mutant and wild-type residues, respectively. Larger values therefore indicate that the mutant is less probable than the wild type under the model. We correlate this score with experimentally measured mutational effects.

For non-canonical amino acids, we evaluate UNAAGI on the DMS benchmark introduced by \citet{rogers2018nonproteinogenic}. The study measured mutational effects for two peptide--protein complexes: PDB ID 5LY1 (JMJD2A/KDM4A bound to the macrocyclic peptide CP2) and PDB ID 2ROC (Mcl-1 bound to PUMA). The dataset reports $\Delta\Delta G$ measurements for substitutions spanning 20 canonical and 20 non-canonical amino acids. To assess performance on canonical amino acids, we use the same 25 ProteinGym assays as in Section~\ref{subsection_rediscover} and evaluate the canonical amino-acid substitutions generated by UNAAGI.\footnote{The computational cost of UNAAGI scales approximately linearly with protein size because it operates on local structural environments. Scaling to larger proteins is therefore not fundamentally prohibitive; however, due to limited computational resources, we evaluate a randomly selected subset rather than the full ProteinGym benchmark.}

 %balancing credibility of results with computational constraints.
%Note that the computational cost of UNAAGI scales linearly with protein size (since only local structural environments are considered), so scaling to larger proteins is not fundamentally problematic. 
%Our model is easily scalable to test over proteins of arbitrary length or size, as we are sampling residues based on local structural environment only. 
% The full list of assays and results is provided in Supplementary \ref{suppl2}.

% To evaluate performance on NCAAs, we use peptide–protein complexes from \citet{rogers2018nonproteinogenic}, specifically PDB ID 5LY1 (JMJD2A/KDM4A bound to a macrocyclic peptide CP2) and PDB ID 2ROC (Mcl-1 bound to PUMA). This dataset includes DMS measurements of \(\Delta \Delta G\) for substitutions across 20 canonical and 20 non-canonical amino acids. For fairness, we restrict evaluation to substitutions where both the mutant and wild-type residues appear in the sampled data.

\subsection{Results}

\subsubsection{Mutational Effect Prediction for Non-Canonical Amino Acids}
\label{NCAA_Benchmarks}

We evaluate predictive performance using the Spearman correlation between mutation scores and experimental $\Delta\Delta G$ values. We first report results on the DMS benchmark containing non-canonical amino acid substitutions. We include PepINVENT, NCFlow, and AlphaFold 3 as baselines. NCFlow was evaluated on the same benchmark in its original study, whereas PepINVENT supports the generation of non-canonical amino acids but was not originally evaluated on DMS data. For PepINVENT, we follow the repository guidelines by masking one residue at a time and generating candidate substitutions at every position in each assay. For peptide--protein complexes, we concatenate the peptide and protein sequences as input to the language model. Sampled SMILES strings corresponding to amino acids in the benchmark are mapped back to residue identities to estimate frequencies. AlphaFold 3 accepts NCAAs with Chemical Component Dictionary (CCD) codes, including those covered by this benchmark; we use the predicted interface TM-score (iPTM) as its mutation score. To ensure consistent directionality across methods, we reverse the signs of the AlphaFold 3 and NCFlow scores. RareFold, by contrast, restricts NCAA inputs to a predefined residue dictionary. Since none of the NCAAs in our benchmark are included in this dictionary, RareFold cannot be evaluated on this benchmark.

\begin{figure}[!ht]
  \centering
  \includegraphics[width=1.0\textwidth]{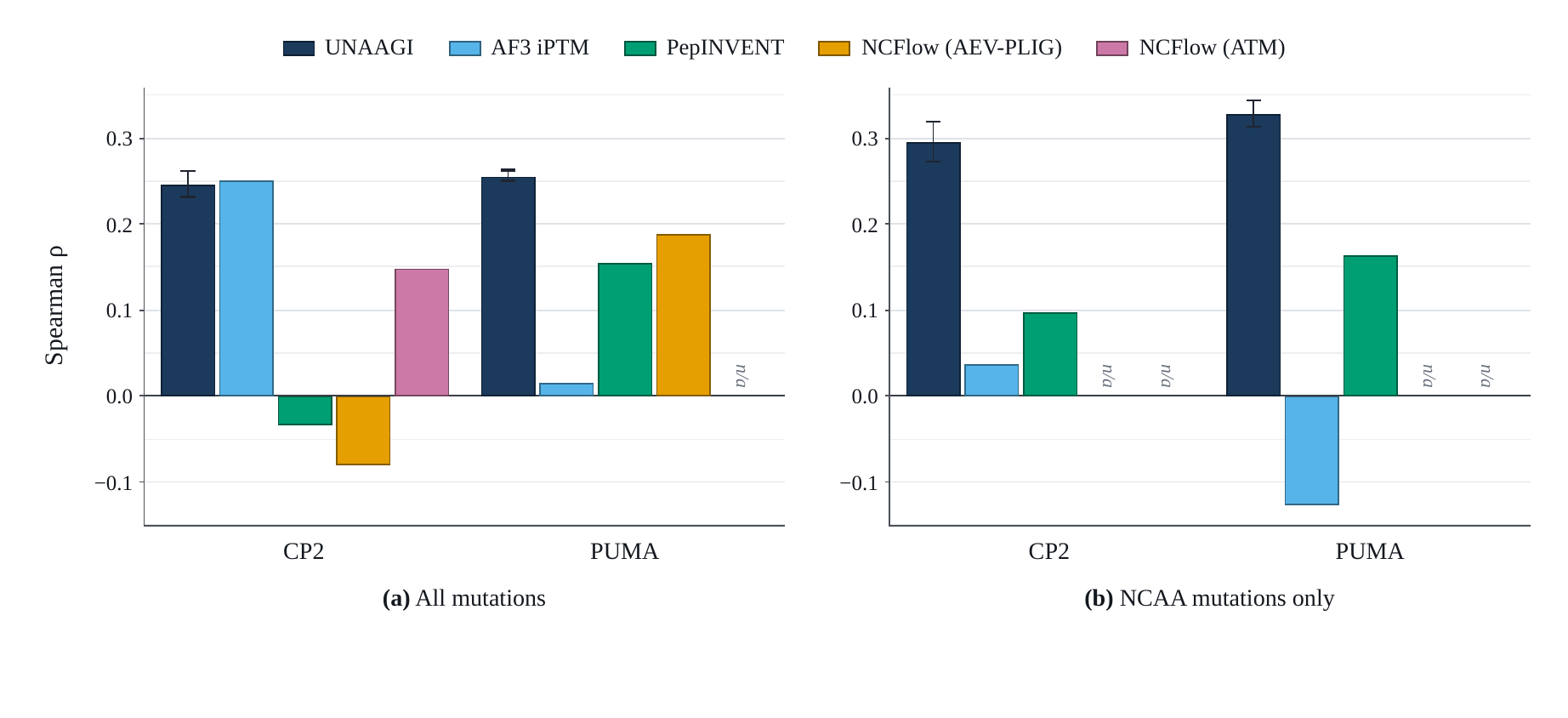}
  \caption{Mutational effect prediction on the CP2 and PUMA DMS benchmarks, evaluated using Spearman correlation. \textbf{(a)} Performance across all measured substitutions, compared with AF3 iPTM, PepINVENT, and NCFlow variants. \textbf{(b)} Performance restricted to non-canonical amino acid substitutions, compared with AF3 iPTM and PepINVENT.}
    \label{vis_ncaa}
\end{figure}

As shown in Figure~\ref{vis_ncaa}, UNAAGI performs consistently across the all-substitution and NCAA-only evaluations. Its correlations on NCAA substitutions are comparable to those obtained when canonical substitutions are also included, suggesting that UNAAGI can generalize beyond 20 canonical amino acids. In contrast, PepINVENT shows limited signal on NCAA substitutions, and NCFlow yields weak or negative correlations regardless of the affinity-prediction module used\footnote{NCFlow results are taken from the original paper. Since NCFlow is not open-sourced, we only report results for all mutations. The original paper reports that the ATM variant fails to produce results for PUMA.}. AlphaFold 3 performs comparably to UNAAGI on CP2 when all substitutions are included, but its performance decreases on NCAA-only substitutions and on PUMA. The PUMA receptor contains less evolutionary information than the CP2 receptor, which may partly explain this performance gap (Appendix~\ref{evo_depth_cp2_puma}). % These results indicate that UNAAGI is the first diffusion-based approach to provide reliable predictive signal on NCAA mutational-effect benchmarks.

%\begin{figure*}[t]
%  \centering
%  \begin{subfigure}[t]{0.49\textwidth}
%    \centering
%    \includegraphics[width=\linewidth]{visualization/barplot_all_mutations.pdf}
%    \caption{All substitutions.}
%    \label{fig:ncaa_all_mutations}
%  \end{subfigure}
%  \hfill
%  \begin{subfigure}[t]{0.49\textwidth}
%    \centering
%    \includegraphics[width=\linewidth]{visualization/barplot_ncaa_mutations.pdf}
%    \caption{Non-canonical substitutions only.}
%    \label{fig:ncaa_only_mutations}
%  \end{subfigure}
%  \caption{Mutational effect prediction on the CP2 and PUMA DMS benchmarks, evaluated using Spearman correlation. \textbf{(a)} Performance across all measured substitutions, compared with AF3 iPTM, PepINVENT, and NCFlow variants. \textbf{(b)} Performance restricted to non-canonical amino acid substitutions, compared with AF3 iPTM and PepINVENT.}
%  \label{vis_ncaa}
%\end{figure*}

\subsubsection{Sanity Check on Canonical Mutational Effect Prediction}
\label{sec:sanity}

As a sanity check, we evaluate UNAAGI on ProteinGym by computing the Spearman correlation between predicted mutation scores and experimental fitness measurements, consistent with the standard metric reported by ProteinGym. Across most assays, UNAAGI exhibits meaningful correlation with experimental fitness (Figure~\ref{vis_proteingym}, Appendix~\ref{app:proteingym}), indicating that its atom-level generative distribution captures biologically relevant variation among canonical amino-acid substitutions. Although UNAAGI is not designed specifically for canonical sequence-based mutational-effect prediction and does not achieve state-of-the-art performance on individual assays, these results provide an independent validation that the structural preferences learned by the model transfer to mutational-effect prediction.

%As shown in Figure~\ref{vis_proteingym},
% Although UNAAGI does not achieve state-of-the-art performance on any individual assay, it exhibits meaningful correlation with experimental mutational effects across most assays (Figure~\ref{vis_proteingym}). Unlike the other baselines, UNAAGI generates residues atom by atom rather than sampling from a fixed vocabulary. ProteinGym is therefore a deliberately challenging evaluation: the possible outputs are known \emph{a priori} to be restricted to the 20 canonical amino acids, whereas UNAAGI uses a more expressive chemical output space. The model nevertheless achieves reasonable predictive performance.

%Furthermore, UNAAGI relies solely on local chemical environments during training and sampling; unlike other baselines, it does not leverage evolutionary information for mutational effect prediction.

% \begin{figure}[h]
%   \centering
%   \includegraphics[width=1.0\textwidth]{visualization/UNAAGI_proteingym_atomic.pdf}
%   \caption{Results of UNAAGI on 20 ProteinGym assays, evaluated using Spearman correlation against experimental DMS measurements. Comparison is made to PepINVENT, as both UNAAGI and PepINVENT generate residues in an atom-wise manner.}
%   \label{vis_proteingym_atomic}
% \end{figure}

Among the baselines, PepINVENT is methodologically closest to UNAAGI because it also represents and samples residues at the atomic level. UNAAGI consistently outperforms PepINVENT across all evaluated assays: PepINVENT shows little or no correlation on most assays, whereas UNAAGI retains substantial predictive signal.\footnote{PepINVENT cannot generate samples when the evaluated sequence exceeds its maximum input length.}

\subsection{Ablation Studies}
\label{ablation}
To assess the contribution of each supplementary training source, we train separate models after removing (i) NCAA-containing PDB structures, (ii) PDBBind complexes, (iii) the independent NCAA dataset, or (iv) all three sources simultaneously. These four ablated models are summarized in Table~\ref{tab:ablation}. UNAAGI (Ablation: NCAA in PDB), UNAAGI (Ablation: PDBBind), and UNAAGI (Ablation: NCAA) denote models trained without the corresponding source; UNAAGI (Ablation: All) is trained only on the canonical PDB dataset.
We evaluate the effect of each ablation by measuring the average Spearman correlation across ProteinGym assays.

\begin{table}[t]
    \centering
    \renewcommand{\arraystretch}{1.3} % increase row height
    \caption{Ablation study of the supplementary training-data sources.}
    \label{tab:ablation}
    \begin{tabular}{lcc}
        \toprule
        \textbf{Model} & \textbf{Spearman $\rho$}  \\
        \midrule
        UNAAGI & \textbf{0.310}  \\
        UNAAGI (Ablation: NCAA in PDB) & 0.267  \\
        UNAAGI (Ablation: PDBBind) & 0.219  \\
        UNAAGI (Ablation: NCAA) & 0.205 \\
        UNAAGI (Ablation: All) & 0.198  \\
        \bottomrule
    \end{tabular}
\end{table}

Removing any supplementary data source reduces performance, indicating that each contributes useful information. The model trained only on the canonical PDB dataset performs worst overall.

\subsection{Capabilities and Limitations}
\label{analysisncaa}

To further investigate the behavior of UNAAGI, we manually inspected a set of sampled amino acids, as shown in Figure~\ref{fig:ncaas}. We observed two main phenomena:
\begin{enumerate}
    \item \textbf{Chemical diversity.} UNAAGI generates amino acids spanning a broad range of chemistries, including hydrophobic, uncharged polar, positively charged, and negatively charged residues.

    \item \textbf{Sampled NCAAs are structurally proximal to canonical amino acids.} UNAAGI is designed to explore non-canonical amino acids in the vicinity of known amino-acid chemistry rather than to generate arbitrarily distant molecular structures. Consistent with this objective, the generated NCAAs often resemble canonical residues or interpolate between them, reflecting both the predominance of canonical amino acids in the training data and the model's focus on chemically proximal substitutions.

\end{enumerate}
In the benchmark described in Section~\ref{NCAA_Benchmarks}, UNAAGI consistently recovers a subset of the evaluated NCAAs: \textbf{Nle}, \textbf{Nva}, \textbf{Abu}, \textbf{Aoc}, \textbf{tBu}, \textbf{hSM}, \textbf{Cpa}, and \textbf{TMe}. However, it does not reliably sample other NCAA classes, including residues with \emph{N}-methyl substitutions or more complex ring systems. Appendix~\ref{supplncaacoverage} provides a more detailed analysis of NCAA coverage. We attribute this limitation in large part to the scarcity of relevant NCAA data. Although consistent with its intended scope, UNAAGI primarily explores NCAAs that are chemically proximal to canonical residues.

% UNAAGI has several important limitations in its current form. Although UNAAGI achieves substantial correlations for the NCAA mutants it successfully samples, coverage remains limited to a subset of the 20 NCAAs in the benchmark (Section~\ref{analysisncaa}). This limitation is driven in large part by the scarcity of relevant NCAA data. Consistent with its intended scope, UNAAGI primarily explores NCAAs that are chemically proximal to canonical residues. Coverage within this space nevertheless remains incomplete, and chemically more distant classes, such as N-methylated residues and complex ring systems, remain challenging.

\section{Conclusion}
We introduced UNAAGI, a molecular diffusion model that samples both canonical and non-canonical amino acids. Across deep mutational scanning benchmarks, UNAAGI achieves meaningful correlations with experimental variant effects without using mutational-effect supervision. Crucially, this predictive signal extends to NCAA substitutions, making UNAAGI, to our knowledge, the first atom-level protein design model to demonstrate positive predictive signal on experimental NCAA mutational-effect benchmarks, while achieving state-of-the-art performance among the evaluated methods.

% atom-level diffusion model to demonstrate positive predictive signal on experimental NCAA mutational-effect benchmarks.

Methodologically, UNAAGI follows the molecular diffusion paradigm developed for structure-based drug design (SBDD), learning a distribution over chemically plausible side chains in protein environments.
%Unlike protein language models \citep{lin2023evolutionary}, UNAAGI does not rely on coevolutionary information from protein sequences, yet still shows predictive ability on mutational effect tasks. 
This formulation highlights a promising connection between SBDD and variant effect prediction, as both depend on modeling local non-covalent interactions. Future work could integrate additional SBDD techniques, such as pharmacophore-based conditioning \citep{Peng2025} or flexible side-chain modeling \citep{schneuing2025multidomain}, to further improve variant effect prediction.

% We observe substantial variability in predictive performance across different assays, a phenomenon shared by all methods. This highlights the intrinsic complexity of the variant effect prediction problem: no single model performs universally well, and outcomes depend strongly on model design, training strategy, and data availability.

Looking forward, several directions could extend UNAAGI. Scaling model capacity and training data may improve coverage, while conditional diffusion could enable classifier-free guidance toward specific NCAAs or desired properties, such as synthetic accessibility.
%  A natural next step is scaling the model, and introduce conditions to it. Incorporating protein structures with NCAA substitutions—although sparse in the PDB—could further expand coverage of the NCAA chemical space. Finally, applying guidance or auxiliary correctors may enhance the generation of chemically diverse NCAAs and enable biasing toward desirable properties, such as synthetic accessibility.

% ----------------------------------------------------------------------------
% Required disclosure statement.
% TODO(author): VERIFY the two factual clauses below before submitting -- what
% AI assistance was actually used, and what it was NOT used for. Only you can
% confirm these, and you take responsibility for the statement.
% ----------------------------------------------------------------------------
\section*{AI use statement}
% \todo[inline]{Verify the scope claims below against your actual tool use.}
We used generative AI tools as general-purpose writing and coding assistants in
the preparation of this work. Specifically, AI assistance was used for
manuscript preparation---copy-editing, \LaTeX{} formatting, and tightening the
exposition of text drafted by the authors---and for writing and debugging
portions of the analysis, evaluation and plotting code supporting the
experiments reported in Section~\ref{experiments}. We did not use generative AI
tools to generate research ideas, to formulate or design the method, to produce
or interpret experimental results, or to retrieve and summarise related work;
the scientific claims, modelling choices and reported numbers in this paper
originate with the authors. All AI-assisted text was reviewed and revised by the
authors, and all AI-assisted code was checked by the authors against the results
it produced. We take responsibility for the final content of this work,
including text, claims and artifacts produced with the aid of generative AI.

% ----------------------------------------------------------------------------
% Recommended disclosure statement.
% ----------------------------------------------------------------------------
\section*{Ethics statement}
This work uses only publicly available structural and mutational-effect data---
the Protein Data Bank, PDBBind, and ProteinGym---and involves no human subjects,
no personally identifiable information, and no newly collected experimental
data. UNAAGI extends structure-based protein design beyond the canonical amino
acid vocabulary, which broadens the accessible chemical space for protein
engineering, therapeutic development and synthetic biology. As with other
generative models in structural biology, this capability carries a dual-use
risk: the same modelling that supports beneficial protein engineering could in
principle be directed toward harmful designs. We regard that risk as limited in
the present setting, since UNAAGI proposes and scores single-residue
substitutions in a given structural context rather than designing complete
functional molecules, and any proposed substitution requires substantial
downstream experimental work before it has a biological effect. We nonetheless
encourage that applications of this and related methods, particularly de novo
design, be pursued with appropriate experimental validation and in adherence to
established biosafety and research-ethics guidelines. The authors declare no
conflicts of interest.

% ----------------------------------------------------------------------------
% Recommended disclosure statement.
\section*{Reproducibility statement}
The information required to reproduce our results is distributed across the main
text and the appendix. The model, its equivariant diffusion formulation and the
training objective are specified in Section~\ref{sec:UNAAGI}, with the full
architecture and hyperparameter settings in Appendix~\ref{suppl1}. Dataset
composition and preprocessing are described in Section~\ref{data_prep}, and the
construction of the NCAA-containing portion of the training data in
Appendix~\ref{train_details}. The sampling and residue-identification procedure
used at inference is given in Section~\ref{sampling}; the held-out amino acid
protocol in Section~\ref{subsection_rediscover}; and the mutational-effect
evaluation, including the DMS benchmarks, in Section~\ref{dms_exp} and
Section~\ref{NCAA_Benchmarks}. Appendix~\ref{eval_clarification} states
precisely what our zero-shot claim does and does not cover, including the
relationship between benchmark proteins and the structural training corpus.
Two sensitivities of the reported numbers are analysed explicitly: robustness to
the pseudocount assigned to unsampled amino acids in
Appendix~\ref{pseudocount_robustness}, and the geometric quality control applied
to generated residues in Appendix~\ref{suppl:qualitative}. All datasets used are
publicly available. We will release all code upon publication.

\bibliography{references}
\bibliographystyle{preprint_style}

\newpage
\appendix

\section{Diffusion and Equivariance Preliminaries}
\label{app:prelim}

This appendix collects the standard formulations referenced in
Section~\ref{sec:UNAAGI}. None of it is specific to UNAAGI.

\subsection{Denoising Diffusion}

Diffusion models have emerged as a powerful class of generative models \citep{sohldickstein2015deepunsupervisedlearningusing, ho2020denoisingdiffusionprobabilisticmodels, song2020generativemodelingestimatinggradients, song2021maximum, song2021scorebasedgenerativemodelingstochastic}. In the discrete-time setting, these models learn to reverse a Markovian forward process that gradually corrupts data with noise until the samples resemble a tractable prior distribution (typically Gaussian). The generative model is trained to approximate the reverse process, which recovers data via a parameterized denoising chain.

Originally developed for image generation, diffusion models have since been extended to molecular and protein structure generation \citep{hoogeboom2022equivariantdiffusionmoleculegeneration}. Let \( \mathbf{x}_0 \sim q(\mathbf{x}_0) \) denote a sample from the data distribution. The forward process is defined as
\begin{equation}
    q(\mathbf{x}_{1:T} \mid \mathbf{x}_0) = \prod_{t=1}^{T} q(\mathbf{x}_t \mid \mathbf{x}_{t-1}),
\end{equation}
which incrementally adds noise over \( T \) steps until \( \mathbf{x}_T \) becomes indistinguishable from Gaussian noise. The reverse process is then modeled as
\begin{equation}
    p_\theta(\mathbf{x}_{0:T}) = p(\mathbf{x}_T) \prod_{t=1}^{T} p_\theta(\mathbf{x}_{t-1} \mid \mathbf{x}_t),
\end{equation}
which gradually denoises \( \mathbf{x}_T \) back to a clean sample.

\subsection{Training Objective}

The training objective follows the standard ELBO on the data log-likelihood:
\begin{equation}
    \log p(\mathbf{x}_0) \geq \mathcal{L}_0 + \mathcal{L}_{\text{prior}} + \sum_{t=1}^{T-1} \mathcal{L}_t,
\end{equation}
where \( \mathcal{L}_0 = \log p(\mathbf{x}_0 \mid \mathbf{x}_1) \) is the reconstruction term and \( \mathcal{L}_{\text{prior}} = -\mathrm{KL}(q(\mathbf{x}_T) \| p(\mathbf{x}_T)) \) matches the prior. In practice, these terms are often omitted. The main learning signal comes from minimizing the per-timestep KL divergence:
\begin{equation}
    \mathcal{L}_t = \mathrm{KL}\!\left[q(\mathbf{x}_{t-1} \mid \mathbf{x}_t, \mathbf{x}_0)\ \|\ p_\theta(\mathbf{x}_{t-1} \mid \mathbf{x}_t)\right],
\end{equation}
which has closed-form solutions under both Gaussian and categorical noise.

\subsection{Equivariance}

Formally, a function \( f: \mathcal{X} \rightarrow \mathcal{Y} \) is equivariant to a group \( G \) if
\[
f(g \cdot x) = g \cdot f(x), \quad \forall g \in G.
\]
For molecular data, the relevant group is \( \text{E}(3) \), which includes rotations, reflections, and translations in 3D. Specifically, a denoising function on atomic coordinates \( \mathbf{X} \in \mathbb{R}^{N \times 3} \) should satisfy
\[
f(\mathbf{X} Q + \mathbf{t}) = f(\mathbf{X}) Q + \mathbf{t},
\]
for any orthogonal transformation \( Q \in \text{O}(3) \) and translation \( \mathbf{t} \in \mathbb{R}^3 \).

\subsection{Message Passing}

Message passing is performed on a fully connected graph with attention-based aggregation. The directional information between atoms \( i \) and \( j \) is encoded as a unit vector
\[
\mathbf{x}_{ji,n} = \frac{\mathbf{x}_j - \mathbf{x}_i}{\|\mathbf{x}_j - \mathbf{x}_i\|},
\]
which is used to construct equivariant vector features. Messages are computed by MLPs over concatenated scalar and edge features and then partitioned into components that update scalar, vector, and positional embeddings equivariantly. This design preserves E(3)-equivariance throughout the network and enables joint modeling of geometry, connectivity, and chemistry during denoising.

\section{Training Details}
\label{suppl1}

\subsection{Architecture Details}
Our model architecture and hyperparameter choices closely follow EQGAT-diff \citep{le2024navigating}. The message-passing function at layer \(l\) is defined as
\[
\mathbf{m}^{(l)}_{ji} = \mathrm{MLP}\!\left(
\left[
\mathbf{h}^{(l)}_{j} \,;\,
\mathbf{h}^{(l)}_{i} \,;\,
\mathbf{W}^{(l)}_{\mathrm{e_0}} \mathbf{e}^{(l)}_{ji} \,;\,
d^{(l)}_{ji} \,;\,
d^{(l)}_{j} \,;\,
d^{(l)}_{i} \,;\,
\mathbf{p}^{(l)}_{j} \cdot \mathbf{p}^{(l)}_{i}
\right]
\right)
\]

where ``;'' denotes concatenation and the MLP is a two-layer perceptron. The resulting message embedding is partitioned into subembeddings:

\[
\mathbf{m}^{(l)}_{ji} = 
\big( 
\mathbf{a}^{(l)}_{ji},\,
\mathbf{b}^{(l)}_{ji},\,
\mathbf{c}^{(l)}_{ji},\,
\mathbf{d}^{(l)}_{ji},\,
\mathbf{s}^{(l)}_{ji}
\big)^\top
\in \mathbb{R}^K
\]

The node, edge, vector, and coordinate updates are then computed as

\[
\mathbf{h}^{(l+1)}_{i}
=
\mathbf{h}^{(l)}_{i}
+
\sum_{j}
\frac{\exp(a^{(l)}_{ji})}{\sum_{j'} \exp(a^{(l)}_{j'i})}
\,\mathbf{W}^{(l)}_{h}\mathbf{h}^{(l)}_{j}
\qquad\text{and}\qquad
\mathbf{e}^{(l+1)}_{ji}
=
\mathbf{W}^{(l)}_{e_1}\sigma(\mathbf{e}^{(l)}_{ji}+\mathbf{d}^{(l)}_{ji}),
\]
\[
\mathbf{v}^{(l+1)}_{i}
=
\mathbf{v}^{(l)}_{i}
+
\frac{1}{N}\sum_{j}
\mathbf{x}_{ji,n}\otimes \mathbf{b}^{(l)}_{ji}
+
\bigl(\mathbf{1}\otimes \mathbf{c}^{(l)}_{ji}\bigr)\odot \mathbf{v}^{(l+1)}_{j}
\,\mathbf{W}^{(l)}_{v},
\]
\[
\mathbf{x}^{(l+1)}_{i}
=
\mathbf{x}^{(l)}_{i}
+
\frac{1}{N}\sum_{j}
s^{(l)}_{ji}\,\mathbf{x}_{ji,n}.
\]

where $\mathbf{1} = (1,1,1)^\top$ and $\sigma$ is the SiLU activation function. After \(L\) layers of message passing, the embeddings are transformed as shown in Figure~\ref{fig:architecture} to obtain the final representations. Categorical atomic features, including hybridization states, formal charges, atom types, aromaticity, ring membership, and atomic degree, are concatenated to form $\hat{H}$. We use the final-layer prediction $\hat{x}_0 = (\hat{\mathbf{X}}, \hat{\mathbf{H}}, \hat{\mathbf{E}})$ for denoising.

\setcounter{figure}{0}
\renewcommand{\thefigure}{A\arabic{figure}}
\begin{figure}
    \centering
    \includegraphics[width=0.5\linewidth]{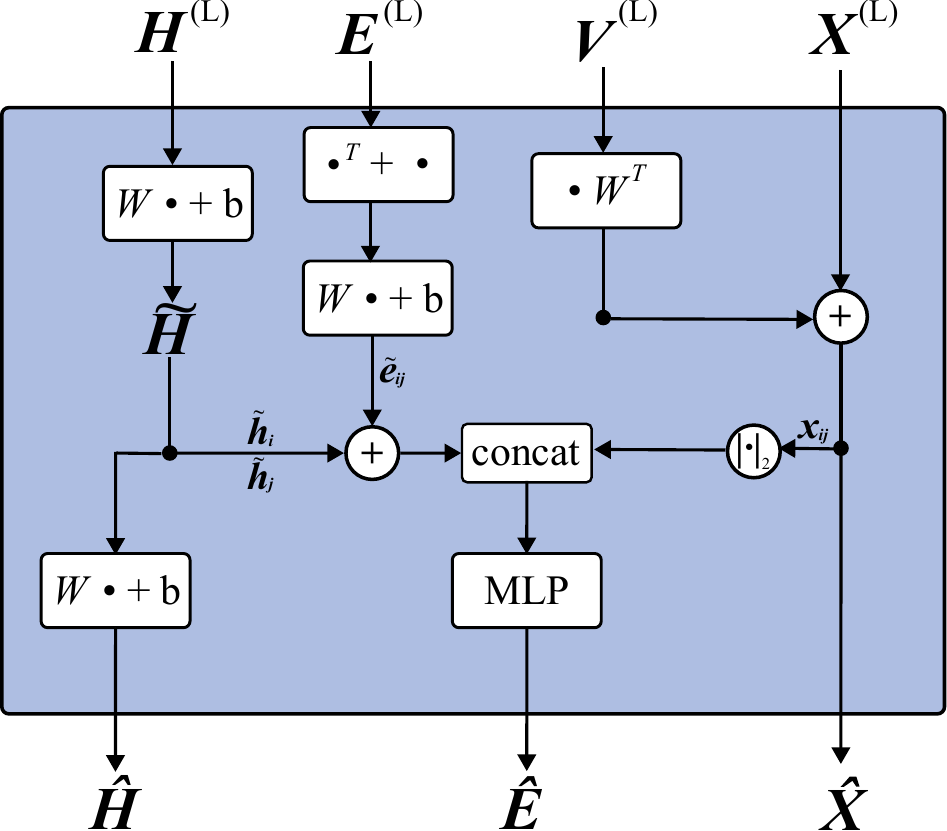}
    \caption{Backbone architecture of UNAAGI, closely following EQGAT-diff \citep{le2024navigating}.}
    \label{fig:architecture}
\end{figure}

\subsection{Hyperparameters}

In all experiments, the forward diffusion process uses 500 timesteps during training. At inference time, we use the 20-step accelerated sampler described in Section~\ref{sampling}. Scalar feature dimensions are set to 256, vector feature dimensions to 64, and edge feature dimensions to 32. We use 7 layers of message passing, resulting in a total of approximately 3.6M trainable parameters.

For optimization, we use Adam with a learning rate of $5 \cdot 10^{-4}$ and weight decay of $1 \cdot 10^{-4}$. Gradients are clipped to a maximum norm of 10. We maintain an exponential moving average (EMA) of the model weights with a decay of 0.9999. The training batch size is 8.

For virtual-node padding, we set $N_{\max}^{\text{train}}=5$ and sample $n_{\text{virt}}\sim U(0,5)$ for each residue during training. Here, $N_{\max}^{\text{train}}$ denotes the maximum number of virtual atoms added to a ground-truth side chain. During sampling, $N_{\max}^{\text{sample}}=15$ denotes the number of nodes in the initial graph. This value covers all amino acids in the benchmark: the largest, 2NP, contains 11 side-chain atoms. During training, side-chain graph sizes range from 0 nodes (Gly with no virtual nodes) to 15 nodes (Trp with five virtual nodes), exposing the model to the graph size used at inference.

\subsection{Details on Training Data Containing NCAAs}
\label{train_details}

We construct the training dataset by filtering PDB entries deposited before January 26, 2026. We retain structures with resolution finer than 2~\AA{}, determined by X-ray crystallography or electron microscopy, and containing full-atom coordinates. We remove hydrogen atoms and isolated water molecules, yielding 102,048 PDB structures for training.

As noted in the dataset composition section, we augment the training set with independent NCAA data from \citet{ilardo2019adaptive} and SwissSidechain \citep{Gfeller2013SwissSidechain}. The dataset of \citet{ilardo2019adaptive} contains 1,913 distinct NCAAs, and SwissSidechain contains 230 NCAAs. These amino acids are used for training as individual molecules without associated structural context.

For the selected PDB structures containing NCAAs, we assemble a dedicated corpus of experimentally determined structures containing NCAAs or stabilizing post-translational modifications (PTMs) \emph{in native structural context}. This dataset, denoted \textsc{NCAA\_PDB}, comprises 14{,}937 high-resolution PDB structures spanning 61{,}937 modified-residue sites.
%, in contrast to the context-free NCAA molecules used during the first training stage. 

%\section{Reconstructing Natural Amino Acids without ever seeing it}

\section{Full ProteinGym Comparison}
\label{app:proteingym}

Figure~\ref{vis_proteingym} reports UNAAGI's Spearman correlation against experimental DMS
measurements across all 25 ProteinGym assays, alongside the baselines discussed in
Section~\ref{dms_exp}. This is the canonical-substitution sanity check described in
Section~\ref{sec:sanity}; UNAAGI is not designed for canonical sequence-based prediction,
and the comparison is included for completeness rather than as a headline result.

\setcounter{figure}{0}
\renewcommand{\thefigure}{C\arabic{figure}}
\begin{figure}[!ht]
  \centering
  \includegraphics[width=1.0\textwidth]{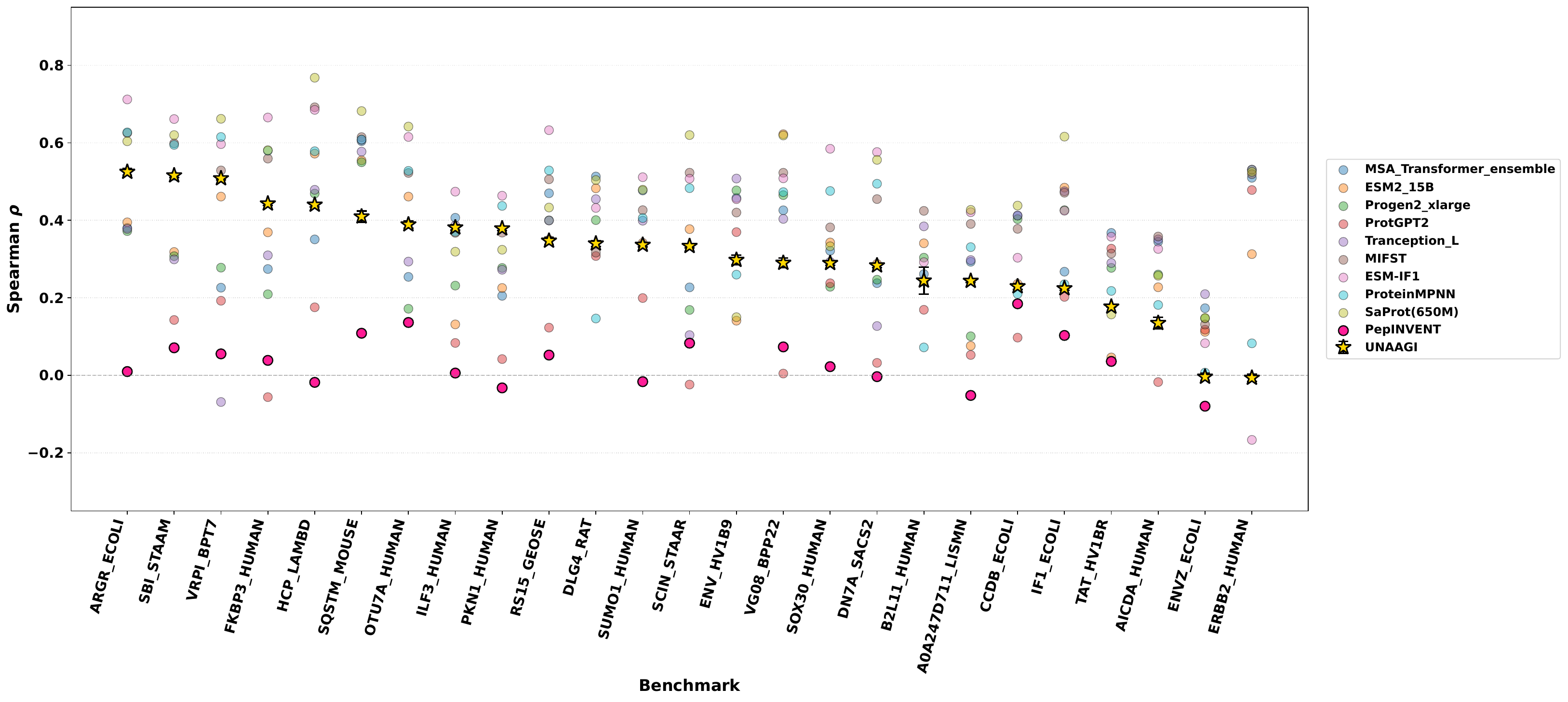}
  \caption{Results on 25 ProteinGym assays, evaluated using the Spearman correlation with experimental DMS measurements. UNAAGI outperforms its closest methodological baseline, PepINVENT, across the evaluated assays.}
    \label{vis_proteingym}
\end{figure}

\section{Additional Robustness Analyses}

\subsection{Held-out Residue Statistics}
\label{AA_recon}

Sample sizes for Table~\ref{tab:heldout-generation}. Reconstruction ($\mathrm{wt}=X$): $n=316$, with 96/91/42/87 sites for VAL/SER/PHE/ILE. Design at favorable sites ($\mathrm{wt}\neq X$): $n=3{,}086$, with 785/704/778/819 sites. Selectivity compares the favorable-site rate ($n=3{,}402$) with the harmful-site rate ($n=1{,}833$, with 434/510/492/397 sites). ``\% kept'' is the ratio between the held-out and full-model rates; the full model achieves $100\%$ reconstruction throughout.

All differences between the held-out and full models at favorable sites are significant (Fisher exact test, $p$: VAL $8.6\mathrm{e}{-6}$, SER $2.1\mathrm{e}{-10}$, PHE $3.0\mathrm{e}{-27}$, ILE $9.2\mathrm{e}{-104}$). Pooled Wilson $95\%$ confidence intervals are $[0.927,0.973]$ for reconstruction, $[0.687,0.719]$ for design, $[0.711,0.741]$ for favorable sites and $[0.599,0.644]$ for harmful sites.

\subsection{Robustness to the Pseudocount for Unsampled Amino Acids}
\label{pseudocount_robustness}

We assess whether the add-one pseudocount assigned to unsampled amino acids drives UNAAGI's ProteinGym performance by recomputing correlations after excluding substitutions that are never sampled. Figure~\ref{fig:proteingym_pseudocount_robustness} compares the original evaluation with this alternative treatment across all 25 ProteinGym assays.

\setcounter{figure}{0}
\renewcommand{\thefigure}{D\arabic{figure}}
\begin{figure*}[t]
    \centering
    \includegraphics[width=\textwidth]{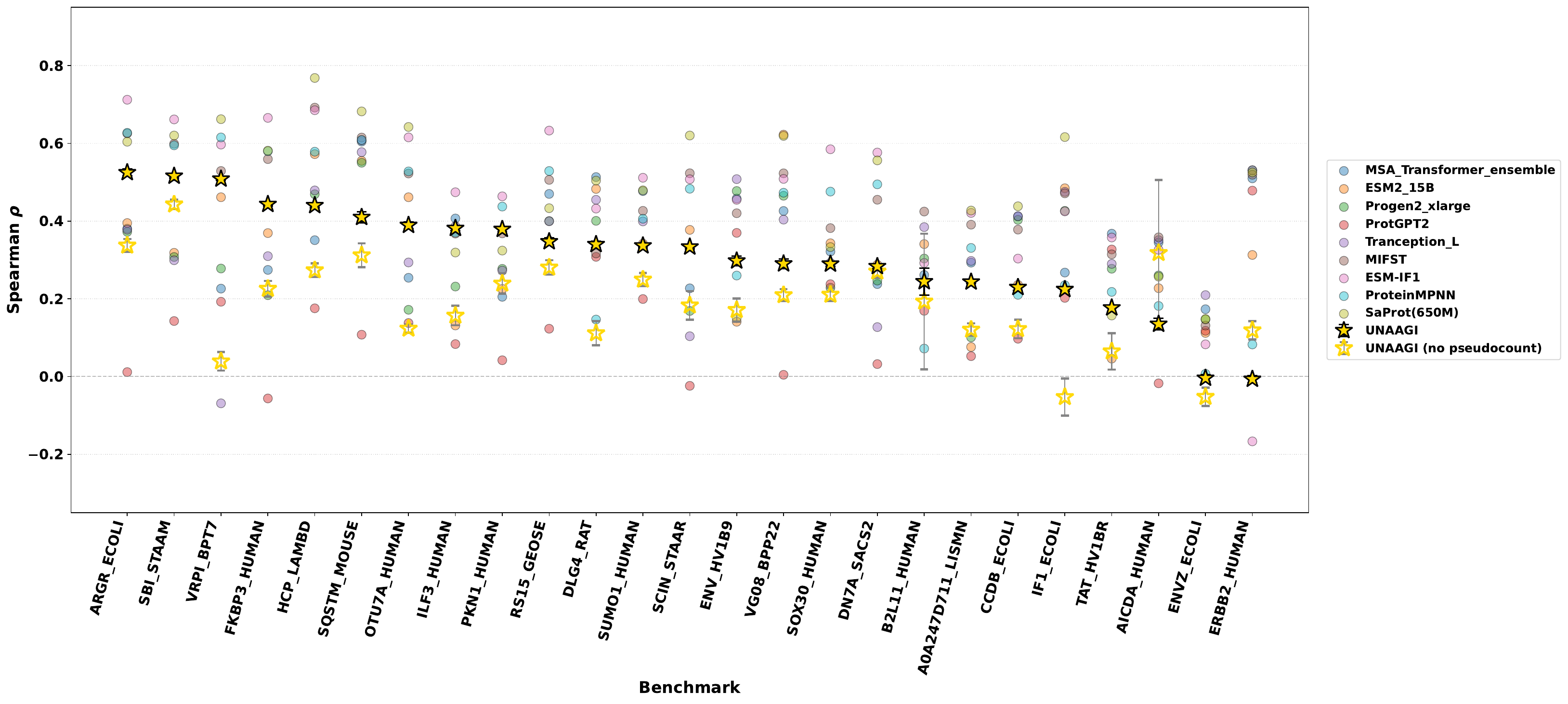}
    \caption{Robustness of ProteinGym performance to the treatment of unsampled amino acids. Spearman correlations are shown for UNAAGI using the add-one pseudocount employed in the main evaluation (black stars) and after excluding unsampled substitutions (yellow outlined stars), alongside sequence- and structure-based baselines.}
    \label{fig:proteingym_pseudocount_robustness}
\end{figure*}

% \subsection{NCAA Coverage Analysis}

\FloatBarrier

\section{Additional Qualitative Results}
\label{suppl:qualitative}

\setcounter{figure}{0}
\renewcommand{\thefigure}{E\arabic{figure}}

\begin{figure}[h!]
    \centering
    \begin{subfigure}[t]{0.48\linewidth}
        \centering
        \includegraphics[width=\linewidth]{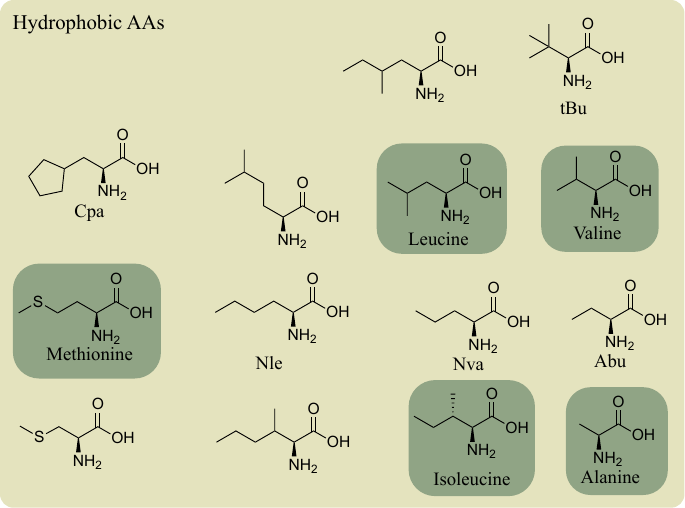}
        \caption{Hydrophobic amino acids sampled by UNAAGI, including both canonical and non-canonical examples (not an exhaustive set).}
        \label{fig:hydrophobic_ncaas}
    \end{subfigure}
    \hfill
    \begin{subfigure}[t]{0.48\linewidth}
        \centering
        \includegraphics[width=\linewidth]{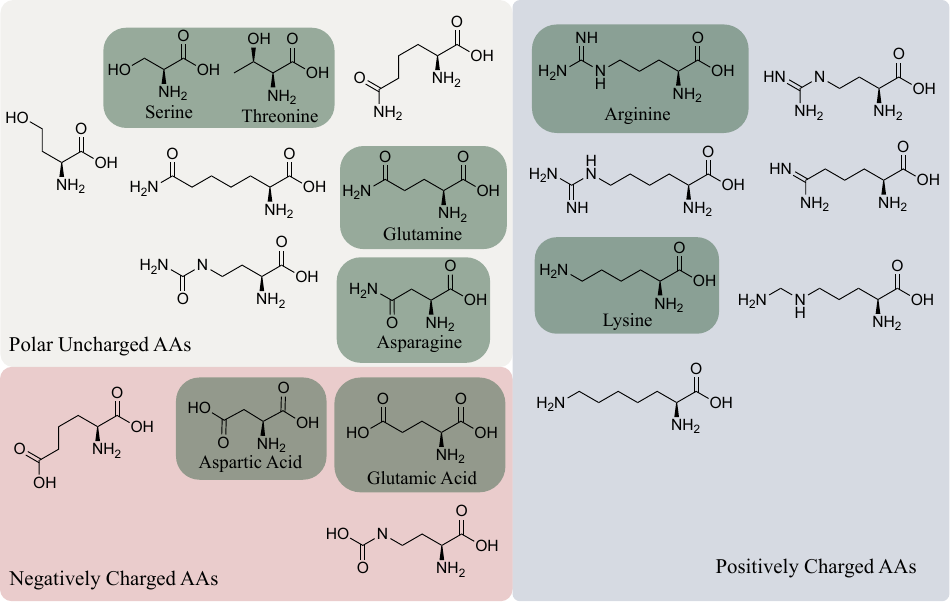}
        \caption{Polar and charged amino acids sampled by UNAAGI, including both canonical and non-canonical examples (not an exhaustive set).}
        \label{fig:other_ncaas}
    \end{subfigure}
    \caption{Qualitative inspection of sampled non-canonical amino acids. Canonical amino acids are highlighted with a green background.}
    \label{fig:ncaas}
\end{figure}

\subsection{Analysis of Wild-Type Coverage}
\label{wt_coverage_supp}

\begin{figure}[h!]
  \centering
  \includegraphics[width=1.0\textwidth]{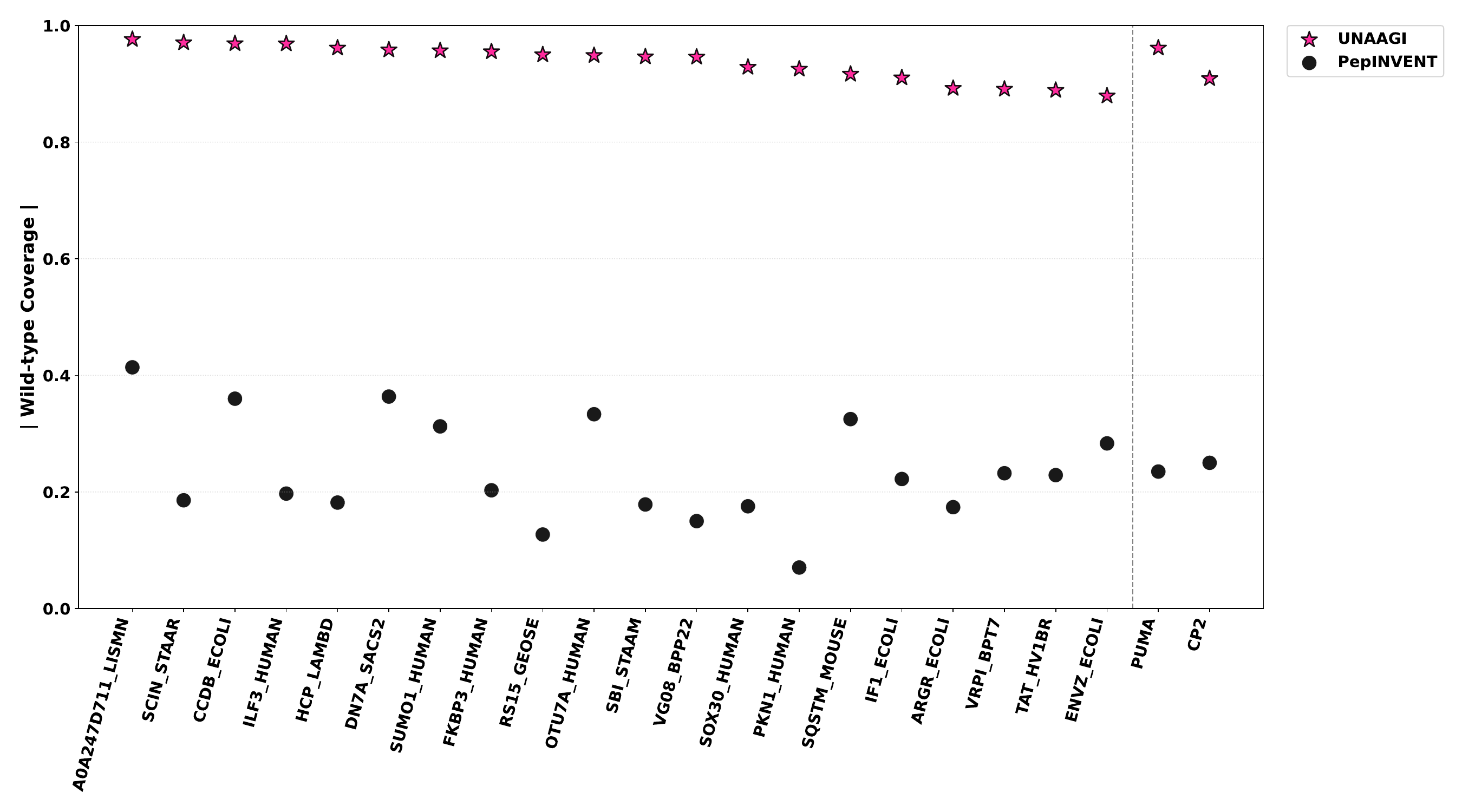}
  \caption{Wild-type coverage rate of UNAAGI and PepINVENT across benchmark assays.}
  \label{vis_coverage_rate}
\end{figure}

Because both UNAAGI and PepINVENT sample residues atom by atom, they must be able to reconstruct wild-type residues reliably. We quantify this ability using the \textit{wild-type coverage rate}, defined as the fraction of positions in an assay for which the wild-type residue is sampled at least once. Table~\ref{tab:wt_coverage} reports the average across all benchmarks, and Figure~\ref{vis_coverage_rate} provides assay-level results. UNAAGI achieves substantially higher coverage, whereas PepINVENT rarely recovers the wild-type residue.

\setcounter{table}{0}
\renewcommand{\thetable}{E\arabic{table}}
\begin{table}[t]
    \centering
        \caption{Comparison of wild-type coverage rates for UNAAGI and PepINVENT, averaged across all benchmarks, including those introduced in Section~\ref{NCAA_Benchmarks}.}
    \label{tab:wt_coverage}
    \renewcommand{\arraystretch}{1.3} % increase row height
    \begin{tabular}{l c}
        \toprule
        \textbf{Model} & \textbf{Wild-Type Coverage Rate} \\
        \midrule
        UNAAGI (ours) & \textbf{0.9368} \\
        PepINVENT     & 0.2365 \\
        \bottomrule
    \end{tabular}
\end{table}

\FloatBarrier

\subsection{Comprehensive Analysis of NCAA Coverage}
\label{supplncaacoverage}

\begin{figure}[h!]
    \centering
    \includegraphics[width=\linewidth]{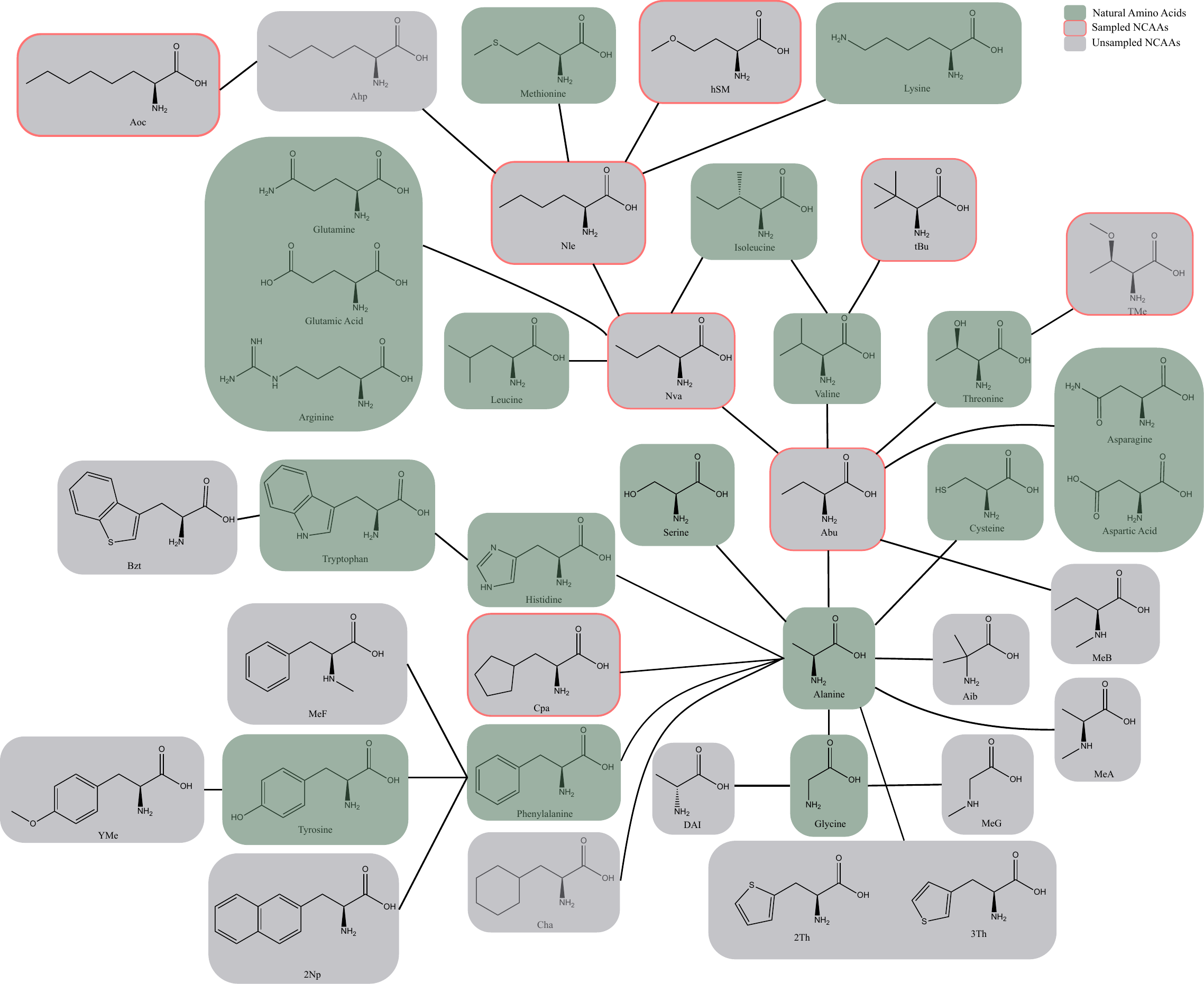}
    \caption{Relationships between canonical amino acids and the NCAAs in the benchmark of \citet{rogers2018nonproteinogenic}. Canonical amino acids are shown with a green background, NCAAs with a gray background, and NCAAs sampled by UNAAGI are highlighted with a red outline. Edges connect amino acids that differ by one graph edit.}
    \label{fig:coverage_graph}
\end{figure}

We examine how UNAAGI's coverage of the NCAAs in the benchmark of \citet{rogers2018nonproteinogenic} relates to their chemical similarity to canonical amino acids. Figure~\ref{fig:coverage_graph} summarizes these relationships as a molecular graph-edit network. The sampled NCAAs tend to be structurally similar to canonical residues, consistent with the model's ability to extend beyond the canonical vocabulary to chemically proximal substitutions. Coverage remains limited for other classes, including \emph{N}-methylated residues and residues with more complex ring systems.

\FloatBarrier

\begin{figure}
    \centering
    \includegraphics[width=\linewidth]{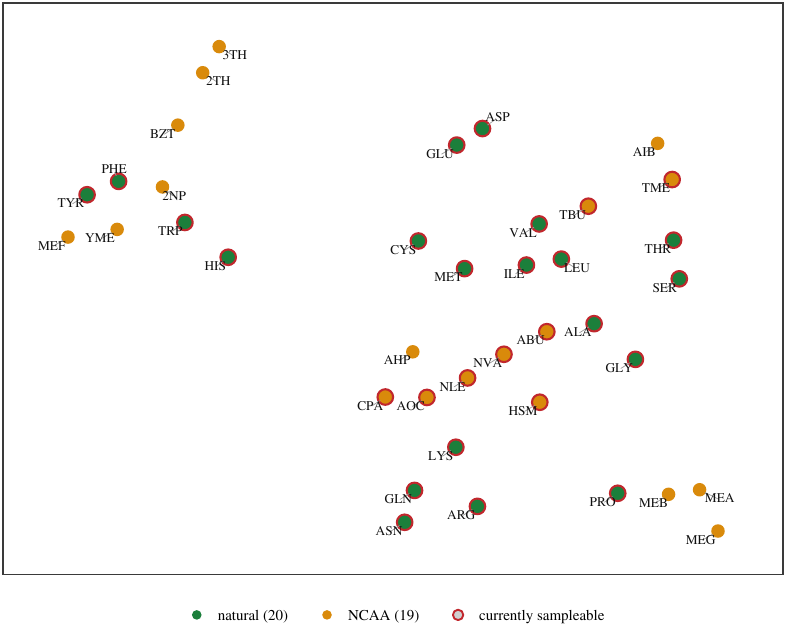}
    \caption{Chemical-space visualization of amino-acid coverage by UNAAGI. Molecular embeddings from the encoder of \citet{rong2020selfsupervisedgraphtransformerlargescale} are projected into two dimensions using UMAP. Each point represents an amino acid in the benchmark: green denotes the 20 canonical amino acids, orange denotes 19 NCAAs, and red outlines indicate amino acids sampled by UNAAGI. Sampled NCAAs are concentrated among residues with aliphatic side chains and related modifications, whereas aromatic and \emph{N}-methylated NCAAs are largely unsampled.}
    \label{fig:ncaa_embedding_coverage}
\end{figure}

To complement this graph-based analysis, we obtain molecular embeddings for the amino acids in the benchmark using the encoder of \citet{rong2020selfsupervisedgraphtransformerlargescale} and project them into two dimensions using UMAP (Figure~\ref{fig:ncaa_embedding_coverage}). The projection reveals chemically related groups, with sampled NCAAs concentrated among residues with aliphatic side chains and related modifications. In contrast, NCAAs containing aromatic rings or \emph{N}-methyl modifications are largely unsampled. This qualitative pattern is consistent with selective coverage of NCAA chemical space, rather than uniform generalization across all chemical classes; the projection itself does not quantify mutational-effect prediction accuracy.

\subsection{Quality Control of Generated Amino Acids with PoseBusters}

\begin{figure*}[t]
    \centering
    \includegraphics[width=\textwidth]{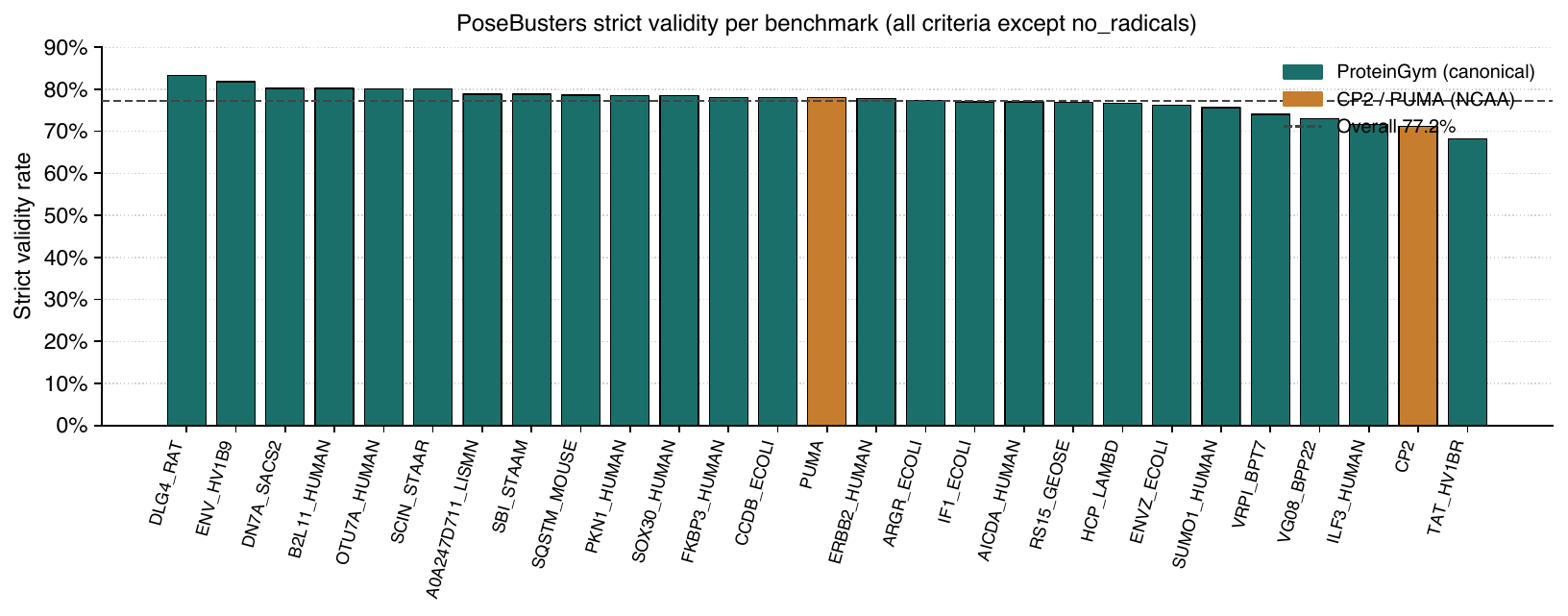}
    \caption{Strict PoseBusters validity rate for each benchmark assay. A generated amino acid is counted as valid only if it passes all included PoseBusters criteria. ProteinGym assays are shown in teal, and the CP2 and PUMA NCAA benchmarks are shown in orange. The dashed line indicates the overall validity rate of 77.2\% across all 13{,}472 samples.}
    \label{fig:posebusters_per_assay}
\end{figure*}

\begin{figure*}[t]
    \centering
    \includegraphics[width=0.75\textwidth]{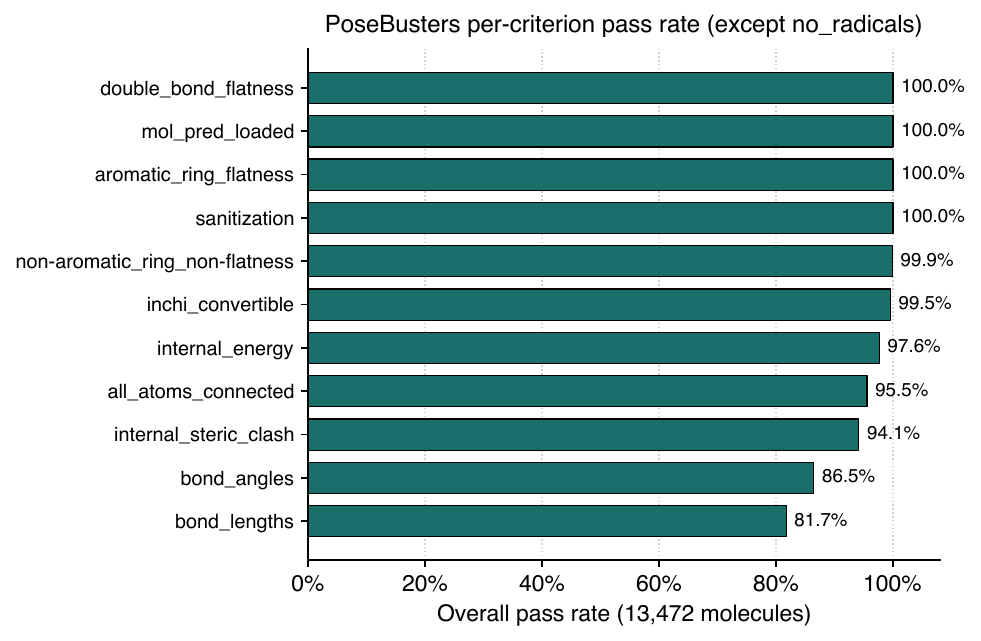}
    \caption{PoseBusters pass rate for each included validity criterion, aggregated across all 13{,}472 generated amino acids. The \texttt{no\_radicals} criterion is excluded. Most samples pass chemical-validity, connectivity, energy, and steric-clash checks, whereas bond-length and bond-angle criteria account for the largest fractions of failures.}
    \label{fig:posebusters_per_criterion}
\end{figure*}

We evaluate the chemical and geometric validity of 13{,}472 sampled amino acids using PoseBusters. A sample is considered strictly valid if it passes every included criterion. Figure~\ref{fig:posebusters_per_assay} reports the strict validity rate for each benchmark, yielding an overall pass rate of 77.2\%. Figure~\ref{fig:posebusters_per_criterion} disaggregates these results by criterion. The generated molecules almost universally pass sanitization, ring-planarity, and molecular-loadability checks, while most failures arise from deviations in bond lengths and bond angles. These results indicate that UNAAGI generally produces chemically coherent molecular graphs, with local geometric accuracy representing the primary remaining source of invalidity.

\FloatBarrier

\subsection{Evolutionary Depth of the CP2 and PUMA Receptors}
\label{evo_depth_cp2_puma}

\begin{figure}[h!]
    \centering
    \includegraphics[width=\linewidth]{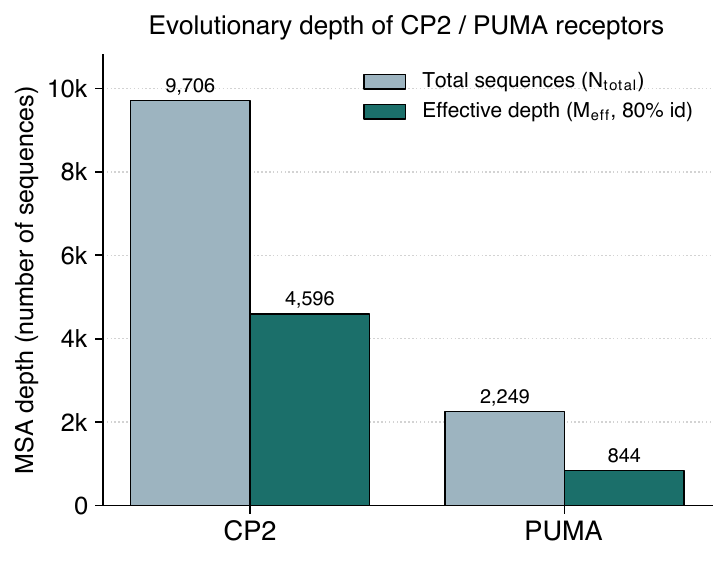}
    \caption{Evolutionary depth of the CP2 and PUMA receptor MSAs. Bars show the total number of sequences ($N_{\mathrm{total}}$) and the effective sequence depth ($M_{\mathrm{eff}}$) after reweighting at an 80\% sequence-identity threshold.}
    \label{fig:cp2puma_evolutionary_depth}
\end{figure}

To characterize the evolutionary information available for the two NCAA mutational-effect benchmarks, we quantify the depth of the multiple-sequence alignments (MSAs) for the CP2 and PUMA receptors. Figure~\ref{fig:cp2puma_evolutionary_depth} reports both the total number of retrieved sequences and the effective depth after sequence reweighting at an 80\% identity threshold. Both receptors have substantial evolutionary coverage, although the CP2 alignment is considerably deeper than the PUMA alignment. This disparity may partly explain the difference in AlphaFold3's mutational-effect prediction performance between the two receptors, as AlphaFold3 uses evolutionary information from MSAs.

\section{Clarification on Evaluation}
\label{eval_clarification}

UNAAGI is evaluated zero-shot with respect to mutational-effect prediction. Training uses molecular structures and residue identities but does not include DMS measurements, $\Delta\Delta G$ values, fitness scores, or task-specific fine-tuning. As with other pretrained zero-shot predictors, benchmark proteins may occur in the structural training corpus. Our evaluation therefore tests whether knowledge learned from structural data without mutational-effect labels transfers to mutational-effect prediction, rather than whether the model generalizes to entirely unseen protein structures.

\end{document}